\documentclass[aps,twocolumn,groupedaddress,floatfix]{revtex4}

\usepackage{epsfig}

\begin{document}

\title{Path-integral Monte-Carlo simulations for electronic dynamics
on molecular chains:\\
II. Transport across impurities}

\author{Lothar M{\"u}hlbacher$^{1}$}
\email[email:~]{lothar.muehlbacher@physik.uni-freiburg.de}
\author{Joachim Ankerhold$^{1,2}$}
\affiliation{1: Physikalisches Institut,
 Albert-Ludwigs-Universit{\"a}t,
 D-79104 Freiburg, Germany}
 \affiliation{2: Service de Physique de l'Etat Condens\'{e},
 Centre d'Etudes de Saclay, 91191\ Gif-sur-Yvette, France}

\date{Date: \today}

\begin{abstract}
Electron transfer (ET) across molecular chains including
an impurity is studied based on a recently improved
real-time path integral Monte Carlo (PIMC) approach [J.
Chem. Phys. {\bf 121}, 12696 (2004)]. The reduced
electronic dynamics is studied for various bridge lengths
and defect site energies. By determining intersite hopping
rates from PIMC simulations up to moderate times, the
relaxation process in the extreme long time limit is
captured within a sequential transfer model. The total
transfer rate is extracted and shown to be enhanced for
certain defect site energies. Further, it is revealed that
the entire bridge compound approaches a steady state on a
much shorter time scale than that related to the total
transfer which allows for a simplified description of ET
along donor-bridge-acceptor systems in the long time range.

\end{abstract}

\maketitle

\newpage

\section{Introduction}

Charge transfer in native or designed molecular structures has attracted a
substantial amount of research recently
\cite{ratner,jortner,dekker,nitzan2,hanggi}. This is mostly due to the
perspective to exploit corresponding processes in a controlled way in
integrated nano-electrical circuits \cite{saclay,weber} . However, even though
tremendous progress has been achieved, still various rather fundamental
questions need to be elucidated. Certainly, one of these is the role of the
environmental degrees of freedom, originating from solvent and/or residual
vibronic molecular degrees of freedom, which are often neglected in approaches
based on transport theories for clean mesoscopic devices. Nevertheless, they
are known to be crucial for charge transfer in condensed phase structures
\cite{marcus56,marcus85,nitzan2}. This is assumed to be particularly true in
cases where impurities along the transport channel are to be surmounted by an
intimate interplay between charge-phonon interactions such as thermal
activation, nuclear tunneling, and coherent electronic tunneling
(superexchange).

The development of effective approaches to include complex environments in the
quantum dynamics of electronic systems has been one of the major goals in
chemistry and physics as well \cite{weiss}. The widely used and powerful
Redfield formulation has revealed important aspects of charge transfer in
various systems \cite{friesner,mukamel,nitzan0,nitzan1}. However, as an
approximate method it is limited to weak friction with high frequency vibronic
modes and sufficiently high temperatures. In a previous paper
\cite{last_paper}, referred to as Ref.~[I] henceforth, we used a numerically
exact Monte Carlo path integral approach (PIMC) to capture electronic transfer
(ET) dynamics in donor-bridge-acceptor (DBA) systems. In essence, it exploits
that the path integral formulation allows for an exact elimination of the bath
as well as the quasi-classical electronic degrees of freedom which reduces the
dimensionality of the Hilbert space considerably \cite{egger_mak}. The method
has the merit of being applicable in the whole parameter space, that is also in
those ranges where approximate methods fail, e.g.\ for low frequency
environments and low temperatures. It further allows to treat bridges with a
larger number of electronic sites and up to sufficiently long times so that in
Ref.~[I] we could investigate the length dependence of the total transfer rate
from donor to acceptor, which has also been the subject of recent experiments
\cite{joachim,ratner_nature,giese}.

In Ref.~[I] we paid particular attention to charge
diffusion across a DBA complex with energetically
degenerated bridge states. Here, we continue this study by
analysing the impact of an impurity located in the center
of the bridge.  One may expect that the existence of a
defect is the typical situation in native degenerated
structures. Moreover, meanwhile advanced chemical
synthesis allow to ``dope'' molecular chains in a
controlled way and study transport properties by linking
them to external leads \cite{weber1}. The archetypical
case is thereby that of a defect with an energy gap to its
surrounding electronic sites as this scenario includes
effectively also defects in the electronic intersite
coupling. Our main focus here lies on a microscopic
description of the transport {\em dynamics} and the total
transfer rate, which for isolated DBA systems corresponds
to the relaxation rate towards thermal equilibrium, and
for integrated DBA structures subject to a dc-voltage is
related to the steady state flux \cite{nitzan2}. In
general, the long time limit where a global thermal
equilibrium state is established, cannot be reached within
typical times accessible in PIMC simulations, however, our
improved method covers time windows from which all
dynamical scales including the relaxation rate can be
extracted. Thus, we use the exact PIMC data to determine
and verify a sequential transfer model and, based on this
procedure, obtain a complete understanding of the
dynamical features of incoherent ET in DBA structures.

The paper is organized as follows. We briefly outline the model for describing
dissipative ET dynamics in Sec.~\ref{Sec:Dynamics of the dissipative d-level
system} and in Sec.~\ref{Sec: Simulation method} the corresponding real-time
PIMC simulation approach; for further details we refer to our previous paper
Ref.~[I]. Then, the numerical results are presented in Sec.~\ref{Sec: Transfer
dynamics}, where we start in Sec.~\ref{Subsec:Local hopping rates} by
extracting intersite hopping rates from the PIMC data based on a sequential
hopping model. The long time dynamics according to these findings is analysed
in Sec.~ \ref{Subsec:Total transfer rate} and the dynamics of the total bridge
compound in Sec.~\ref{Subsec:Bridge relaxation}. At the end a short summary and
some conclusions are given.

\section{Dynamics of the dissipative $d$-level system}
\label{Sec:Dynamics of the dissipative d-level system}

\begin{figure}
\epsfxsize=8cm \epsffile{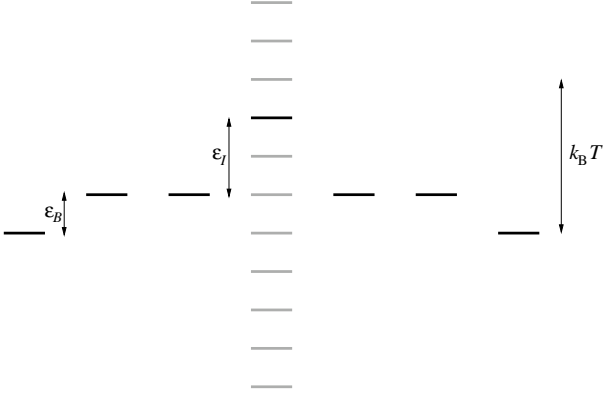} \caption[]{\label{fig1}
Molecular chain with $b = 5$ bridge states ($d=7$) and an impurity in its
center. Gray bars denote the different energetic layouts for the impurity
investigated in this paper.}
\end{figure}

We investigate the dynamics of a single electron moving on
an impurity flawed molecular chain. The general approach
to describe single charge transfer across bridges has been
outlined in Ref.~[I]. Here, we thus collect only the key
ingredients. The molecular chain consists of $d=2 S+1$
discrete sites (see Fig.~\ref{fig1}), where, separated by
equal distances $a$, $b = d-2$ adjacent bridge states
$\{|B1\rangle, \ldots, |B_b\rangle\}$ are clamped between
a donor $|D\rangle$ and an acceptor state $|A\rangle$. The
latter two are energetically degenerated and define the
$0$-energy level. The bare bridge states are also
energetically degenerated but at a higher energy
$\hbar\epsilon_{B_i} = \hbar\epsilon_B = 2.5\hbar\Delta$,
while the impurity, located in the center of the bridge
$|I\rangle = |B_{(d+1)/2}\rangle$, exhibits an energy gap
$\epsilon_I = \epsilon_{B_{(d+1)/2}} - \epsilon_B$ with
respect to the other bridge states (we only regard chains
symmetric with respect to the impurity site, i.e. with an
odd number of bridge sites). The focus of our study lies
then on the transport across the bridge/impurity compound
when this gap is energetically raised or lowered.

We restrict our model to nearest-neighbor coupling, that
is, electronic motion is facilitated through tunneling
which only occurs between adjacent states with a uniform
real tunneling amplitude $\Delta_{s,s'}$. The electronic
coordinate can then be expressed as
\begin{equation} \label{path}
q(t) = a \cdot s(t) \;,
\end{equation}
where $-S \le s(t) \le S$. The position operator thus is equivalent to the spin
$S$ operator
\begin{equation}
a{\mathbf S}_z|s\rangle = as|s\rangle \;,
\end{equation}
with $|s\rangle$ denoting the (orthonormal) localized electronic states. In
terms of electron transfer, $|{-S}\rangle$ and $|S\rangle$ represent the donor
and acceptor $|D\rangle$ and $|A\rangle$, respectively, while the other states
are referred to as the bridge states.

The free $d$-level system ($d$LS) Hamiltonian can then be
written as Ref.~[I].
\begin{equation} \label{H_dLS-Hamiltonian}
H_{d\rm LS} = \hbar {\mathbf E}_z - \hbar {\mathbf S}_x \;,
\end{equation}
where ${\mathbf E}_z$ describes the energetic distribution of the electronic
sites according to
\begin{eqnarray}
{\mathbf E}_z|D\rangle  &=& {\mathbf E}_z|A\rangle = 0 \;, \nonumber\\
{\mathbf E}_z|B_i\rangle &=& \epsilon_B|B_i\rangle \quad {\rm for}\ 1 \le i \le b
\;,\ i \neq (d+1)/2 \;, \nonumber\\
{\mathbf E}_z|I\rangle  &=& (\epsilon_B + \epsilon_I)|I\rangle \;,
\end{eqnarray}
while ${\mathbf S}_x$ governs the tunneling,
\begin{eqnarray}
{\mathbf S}_x|D\rangle &=& \Delta|B_1\rangle \;, \nonumber\\
{\mathbf S}_x|B_1\rangle &=& \Delta(|D\rangle + |B_2\rangle) \;, \nonumber\\
{\mathbf S}_x|B_{(d+1)/2\mp 1}\rangle &=&
\Delta(|B_{(d+1)/2\mp 2}\rangle
                            + |I\rangle) \;, \nonumber\\
{\mathbf S}_x|I\rangle &=& \Delta(|B_{(d+1)/2-1}\rangle \nonumber\\
        && \quad{} + |B_{(d+1)/2+1}\rangle) \;, \nonumber\\
{\mathbf S}_x|B_i\rangle &=& \Delta(|B_{i-1}\rangle + |B_{i+1}\rangle) \nonumber\\
        && {\rm for}\ i \in \{2, \ldots (d+1)/2-1, \nonumber\\
        && \quad (d+1)/2+1, \ldots, d-1\} \;, \nonumber\\
        {\mathbf S}_x|B_b\rangle &=& \Delta(|A\rangle + |B_{b-1}\rangle) \;, \nonumber\\
{\mathbf S}_x|A\rangle &=& \Delta|B_b\rangle \;, \nonumber\\
\end{eqnarray}
The interaction with solvent and vibronic degrees of freedom is introduced
within the framework of a system-plus-reservoir model, leading to the total
Hamiltonian \cite{weiss}
\begin{eqnarray} \label{hamilton}
H &=& H_{d\rm LS} + H_I + H_B \nonumber\\
&=& H_{d\rm LS}- a{\mathbf S}_z \sum_\alpha c_\alpha X_\alpha
 + (a{\mathbf S}_z)^2 \sum_\alpha {c_\alpha^2 \over 2m_\alpha \omega_\alpha^2}
\nonumber\\
&&{} + \sum_\alpha \left( {P_\alpha^2 \over 2m_\alpha}
 + {1\over2}m_\alpha\omega_\alpha^2 X_\alpha^2 \right) \; .
\end{eqnarray}
Here, the residual degrees of freedom are archetypically
modeled as a harmonic bath coupling bilinearly  to the
position of the electron ($H_I$) \cite{weiss}. As
discussed in detail in Refs.~\cite{chandler,song,weiss},
this provides a reasonably accurate description of reality
for the great majority of ET systems.

For the dissipative electronic dynamics we focus on the {\it time-dependent
site populations},
\begin{equation} \label{populations}
P_{s_f,s_i}(t) = {\rm Tr}\left\{e^{iHt/\hbar} |s_f\rangle\!\langle s_f|
e^{-iHt/\hbar} W_i(0)\right\} \;,
\end{equation}
which are normalized, $\sum_{s_f=-S}^S P_{s_f,s_i}(t) = 1$, and where $W_i(0)$
specifies the initial state of the total compound. In most theoretical and
experimental works an initial separation of the electron and the environment is
assumed corresponding to an initial density matrix
\begin{equation} \label{initial_perp_P}
W_i(0) = Z_B^{-1} |s_i\rangle\!\langle s_i| e^{-\beta(H_B-s_i \mu{\cal
E})} \;,
\end{equation}
with the electron held fixed in state $|s_i\rangle$ {and the bath normalization
$Z_B = {\rm Tr}\{e^{-\beta H_B}\}$ assuring} {the full system's density matrix
to be normalized for all times}. For a transfer process across the entire chain
one typically prepares the electron initially in the donor state, i.e.\
$|s_i\rangle=|D\rangle$ \cite{friesner,nitzan1}. Above, $\mu$ is the dipole
moment of the electron, and ${\cal E}$ denotes the dynamical polarization of
the bath \cite{weiss}, which is equilibrated with respect to the initial
position of the electron. By comparing with Eq.~(\ref{hamilton}), we see that
$\mu {\cal E} = a\sum_\alpha c_\alpha X_\alpha$. As pointed out in
Ref.~\cite{mlb03}, this ``standard preparation'' often used in ET experiments
is especially suitable for a theoretical description of thermal transfer rates.

Switching to the path integral representation of Eq.~(\ref{populations}), the
bath degrees of freedom can be traced out exactly, yielding the reduced
dynamics \cite{weiss}
\begin{equation} \label{populations path-integral}
P_{s_f,s_i}(t) = \frac{1}{Z} \oint\!{\cal D}\tilde{s}\;
 \delta_{\tilde{s}(t),s_f} \exp\left\{ {i\over\hbar}S_{d\rm LS}[\tilde{s}] -
 \Phi[\tilde{s}] \right\} \;.
\end{equation}
Here the path integration runs over closed paths $\tilde{s}(\tilde{t})$
starting at $\tilde{s}(0)=s_i$ and propagating along the real-time contour
$\tilde{t} \in 0 \rightarrow t \rightarrow 0$, which connect the forward and
backward paths $s(t')$ and $s'(t')$, respectively. Furthermore, $S_{d\rm
LS}[s]$ denotes the total action of the free $d$LS. The influence of the
traced-out bath is completely encoded in the {\sl Feynman-Vernon influence
functional} $\Phi[s]$ \cite{feynman},
\begin{eqnarray} \label{influence-exponent}
\Phi[s, s'] &=& \int_0^{t}\!dt'\! \int_0^{t'}\!dt'' [s(t') - s'(t')]
\ [{L}(t'-t'')s(t'') \nonumber\\
&& \qquad {}- {L}^\ast(t'-t'')s'(t'')] \nonumber\\
&& {}+ i{{\hat{\mu}} \over 2}\int_0^{t}\!dt' [s^2(t') - s'^2(t')] \;,
\end{eqnarray}
which is written in terms of the complex-valued bath
autocorrelation function reading for real time $t$
\begin{eqnarray} \label{L(t)}
L(t) &=&\frac{1}{\hbar^2}\left\langle \left(\sum_\alpha c_\alpha
X_\alpha(t)\right)\left(\sum_\alpha c_\alpha X_\alpha(0)\right) \right\rangle_\beta\nonumber\\
&=& {1\over\pi}\int_0^\infty \!d\omega\, J(\omega)
{\cosh[\omega(\hbar\beta/2-it)] \over
\sinh(\hbar\beta\omega/2)}
\end{eqnarray}
with $\beta=1/k_B T$ and
\begin{equation} \label{muu}
{\hat{\mu}} = {2 \over \pi} \int_0^\infty\!d\omega {J(\omega) \over
\omega} = 2\Lambda_{\rm cl} \;.
\end{equation}
Both quantities are completely determined by the spectral density
\begin{equation}
J(\omega) = \frac{\pi a^2}{2\hbar} \sum_\alpha \frac{c_\alpha^2}
 {m_\alpha\omega_\alpha} \delta(\omega-\omega_\alpha) \;,
\end{equation}
which for a condensed-phase environment typically becomes a continuous function
of $\omega$. In the classical regime, the description of the bath reduces to
the celebrated Marcus theory \cite{marcus56,marcus85}, where the dissipative
influence onto the electronic dynamics are fully governed by the classical {\sl
reorganization energy} $\hbar \Lambda_{\rm cl}$.

\section{Simulation method}
\label{Sec: Simulation method}

We perform path-integral Monte Carlo (PIMC) simulations, utilizing a
discretized version of the path-integral expression (\ref{populations
path-integral}) where the electronic system's path $s(t)$ is replaced by
$\{s_j\}_{1 \le j \le 2q}$, $s_j = s((j-1)t/q)$. The ``dynamical sign problem''
\cite{sign-problem} notorious for real-time calculations of quantum systems is
significantly relieved by employing an optimized filter based on the blocking
approach which exploits certain symmetries of the influence functional
\cite{egger_mak}. Besides some improvements described below, we used the same
computer code as described in detail in Ref.~[I].

The computational bottleneck of this approach is the creation of the MC
trajectory since the evaluation of each trial move requires the calculation of
a series of matrix multiplications. However, here significant saving can be
achieved. Starting with the standard MC weight,
\begin{eqnarray} \label{MC-weight 1}
\lefteqn{w_{\rm MC}(\{\xi_j\}, \eta_{q+1}) \equiv
 \exp\!\Bigg(-\sum_{k \ge j = 2}^q \xi_k \Lambda_{k-j}
 \xi_j \Bigg)} \nonumber\\
&& \times \left|\left(\hat{K}^{(1)}(\xi_2,\dots,\xi_q) \cdots
 \hat{K}^{(q)}(\xi_q)\right)_{2s_i,\eta_{q+1}}\right| \;,
\end{eqnarray}
where the exponential is the real part of the influence functional while the
second factor represents a product of complex matrices, which include the
complex valued contributions from the influence functional and the propagation
of the undamped electronic system (for details, see~Ref.~[I] App.~A), in a
first step, these matrices are stripped off all dissipative terms leading to
\begin{eqnarray} \label{MC-weight 2a}
\tilde{w}_{\rm MC}(\{\xi_j\}, \eta_{q+1}) &\equiv&
 \exp\!\Bigg(-\sum_{k \ge j = 2}^q \xi_k \Lambda_{k-j}
 \xi_j \Bigg) \\
&& \times \left|\left(\tilde{K}(0,\xi_2) \cdots
 \tilde{K}(\xi_q,0)\right)_{2s_i,\eta_{q+1}}\right| \,. \nonumber
\end{eqnarray}
Now the matrices $\tilde{K}(\xi_j,\xi_{j+1})$ can be computed from the free
propagator alone and thus are, unlike the $\hat{K}^{(j)}$ matrices in
Eq.~(\ref{MC-weight 1}), {\sl local in time} in the sense that they connect
only two neighboring time slices. Accordingly, the former can be calculated and
stored beforehand for all combinations of $\xi_j$ and $\xi_{j+1}$, while the
latter have to be recalculated every time a trial move is evaluated.

The respective speed up can be multiplied by computing and storing sequences of
$\tilde{K}$ matrices prior to performing the MC moves, e.g. with
\begin{equation} \label{Rest vom Fest}
\tilde{K}^{(n)}(\xi_1, \dots, \xi_{n+1}) \equiv
\prod_{j=1}^n \tilde{K}(\xi_j, \xi_{j+1}) \;,
\end{equation}
which diminishes the number of matrix multiplications necessary to obtain
$\tilde{w}_{\rm MC}$ by a factor of $n$. Of course this approach is limited by
its memory consumption which increases exponentially with $d$.  However, with
the on-board memory facilities of nowadays computer's being in the GByte range,
it still offers tremendous performance improvements in terms of simulation
duration.

Finally, in a last step the $\tilde{K}^{(n)}$ are replaced
by real-valued matrices which are obtained from the former
by replacing all matrix entries with their modulus.
Similarly to ignoring the dissipative terms present in the
original $\hat{K}^{(j)}$ matrices, this somewhat impairs
the statistics of the PIMC simulation. However, in both
cases the thus necessarily larger number of accumulation
processes is more than compensated by the speed up of the
evaluation of the trial moves, provided the system is not
too far from the coherent-incoherent transition.
Fortunately, far away from this transition the application
of special techniques to soothe the sign problem usually
is not necessary anyway. Note also that the deviation of
the improved MC weight~(\ref{MC-weight 2a}) from
Eq.~(\ref{MC-weight 1}) decreases with increasing $n$.

To illustrate the effectiveness of this approach we
performed calculations both with the improved as well as
with the original MC weight. For the parameters used in
this paper, the former achieved a speed up of typically a
factor of 12; the data for $\epsilon_I/\Delta = 5$
presented in Fig.~\ref{fig2a}, e.g., were obtained on
IBM-P615 within a CPU time of 6.2 days with $n=4$
(cf.~Eq.~(\ref{Rest vom Fest}) and $q=100$ discretization
steps on the real-time axis, where $6.7 \times 10^7$
accumulation processes were performed; the statistical
error in the average sign was $1.2\%$.

\section{Transfer dynamics}
\label{Sec: Transfer dynamics}

For vanishing dissipation, the charge dynamics of an electron initially
localized at the donor site displays coherent oscillations corresponding to no
net-transfer to the acceptor site. These oscillations survive for weak coupling
to an environment and sufficiently low temperatures, but with decreasing
amplitude as the system relaxes towards its equilibrium state. In the sequel,
we are interested in completely incoherent transport, where, however, quantum
mechanical properties of the phonon bath play a substantial role and give rise
e.g.\ to nuclear tunneling. According to our previous study, we consider a
spectral density of the form
\begin{equation}
J(\omega)=2 \pi \alpha\omega {\rm e}^{-\omega/\omega_c},
\end{equation}
where the damping strength is chosen as $\alpha=1/4$, well below the
coherent-incoherent transition $\alpha\approx 1/2$, and the cut-off frequency
$\omega_c/\Delta=5$. As seen in Ref.~[I], for these parameters coherent
oscillations in the electronic dynamics show up for $\hbar\beta\Delta >
0.3$. Therefore we fix the inverse temperature sufficiently above this
transition at $\hbar\beta\Delta = 0.1$, which gives still rise to strong
quantum effects in the bath ($\omega_c\hbar\beta=0.5$) though.

Following these lines, we expect an initial nonequilibrium state to decay --
after an initial transient on a time scale $\tau_{\rm trans}$ --
multi-exponentially towards thermal equilibrium.  According to our previous
simulations [I], the equilibrium occupation probabilities are taken to be
Boltzmann distributed, i.e.,
\begin{eqnarray}
P_D^\infty &=& P_A^\infty = \frac{1}{2 + (b-1)
e^{-\hbar\beta\epsilon_B} +
 e^{-\hbar\beta(\epsilon_B + \epsilon_I)}} \nonumber\\
&=& \frac{e^{\hbar\beta\epsilon_B}}{b-1 +
2e^{\hbar\beta\epsilon_B} +
 e^{-\hbar\beta\epsilon_I}} \;, \nonumber\\
P_{B_i}^\infty &=& \frac{1}{b-1 + 2e^{\hbar\beta\epsilon_B}
+
 e^{-\hbar\beta\epsilon_I}} \;, \nonumber\\
P_I^\infty &=& \frac{e^{-\hbar\beta\epsilon_I}}{b-1 +
2e^{\hbar\beta\epsilon_B} +
 e^{-\hbar\beta\epsilon_I}} \;.
 \label{equipops}
\end{eqnarray}
Here, $P_D^\infty, P_A^\infty, P_I^\infty$, and $P_{B_i}^\infty$, $i=1, \ldots
(d+1)/2-1,(d+1)/2+1, \ldots, b$ denote the populations of donor, acceptor,
impurity, and the remaining bridge sites, respectively, for $t\to \infty$. For
intermediate to long times this decay is governed by a mono-exponential
relaxation with a time constant that is the inverse of the total transfer rate
$\Gamma_T$. The crucial question is then, to what extent signatures of the
impurity can be detected in $\Gamma_T$, which often is the only observable
accessible in experiments. In case that the molecular chain is connected to
external leads being integrated into an electrical circuit, an applied
dc-voltage gives rise to a steady state with a flux that is directly related to
$\Gamma_T$ \cite{nitzan2}.

\subsection{Population dynamics and intersite hopping rates}
\label{Subsec:Local hopping rates}

For molecular chains with energetically degenerated bridge states, but variable
length treated in Ref.~[I], a sequential hopping model turned out to capture
the population dynamics for times beyond the transient time scale quite
well. Further, direct transfer populating bridge sites only virtually, known as
superexchange, turned out to play no substantial role for the bath parameters
specified above even for larger energy gaps between the intermediate and
donor/acceptor sites.  Consequently, we assume first a purely sequential model
to be applicable also here which is self-consistently verified by analysing the
QMC data. The influence of superexchange is included in a second step.

Then, the population dynamics for ${\mathbf P}=(P_D, P_{B_1}, \ldots, P_{B_b},
P_A)$ is governed by
\begin{equation} \label{differential rate equation}
\dot{{\mathbf P}} = A {\mathbf P} \;,
\end{equation}
with a rate matrix independent of time for $t>\tau_{\rm
trans}$
\begin{widetext}
$A =$ \hspace*{\fill}
\begin{equation}
\left( \label{microscopic rate matrix}
\begin{array}{ccccccccccccccc} 
-\Gamma_{DB}& \Gamma_{BD}         &      & 0        &      &&&&&&& \cdots &&& 0 \\
\Gamma_{DB} &-\Gamma_{BD}-\Gamma_B&      & \Gamma_B &      & 0 &&&&&& \cdots &&& \\
0           & \Gamma_B            &      &-2\Gamma_B&      & \Gamma_B  &      & 0 &&&&\cdots &&& \vdots \\
\vdots      &                     &\ddots&          &\ddots&           &\ddots&&&&&&&& \\
            & 0                   &      & \Gamma_B &      &-\Gamma_B-\Gamma_{BI} &      & \Gamma_{IB} & 0 &&&&&& \\
            &                     &      & 0        &      &\Gamma_{BI}&      &-2\Gamma_{IB}&      &\Gamma_{BI}& 0    &          &      &                     &            \\
            &                     &      &          &      & 0         &      & \Gamma_{IB} &      & -\Gamma_{BI}-\Gamma_B&      & \Gamma_B & 0    &                     &            \\
            &                     &      &          &      &           &      &             &\ddots&           &\ddots&          &\ddots&                     & \vdots     \\
\vdots      & \cdots              &      &          &      &           &      & 0           &      & \Gamma_B  &      &-2\Gamma_B&      & \Gamma_B            & 0          \\
            &                     &      & \cdots   &      &           &      &             &\ddots& 0         &      & \Gamma_B &      &-\Gamma_B-\Gamma_{BA}& \Gamma_{AB}\\
0           & \multicolumn{10}{c}{\cdots}                                                                             & 0        &      & \Gamma_{BA}         &-\Gamma_{AB}\\
\end{array}
\right)
\end{equation}
\end{widetext}
Note that Eq.~(\ref{differential rate equation}) can be derived from an exact
non-local retarded Master equation \cite{nica}. Due to the sequential hopping,
the intersite rates $\Gamma_{DB}$, $\Gamma_{BD} = (P_B^\infty/P_D^\infty)
\Gamma_{DB}$, and $\Gamma_B$ should be independent of the gap energy
$\hbar\epsilon_I$ of the impurity, as our simulations confirm (see below). The
latter one determines only $\Gamma_{BI}$ and $\Gamma_{IB}$ which are related by
the detailed balance condition
\begin{equation} \label{detailed-balance impurity rate}
\Gamma_{BI} = {P_B^\infty \over P_I^\infty} \Gamma_{IB} =
e^{\pm\hbar\beta\epsilon_I} \Gamma_{IB} \; .
\end{equation}
%
\begin{figure}
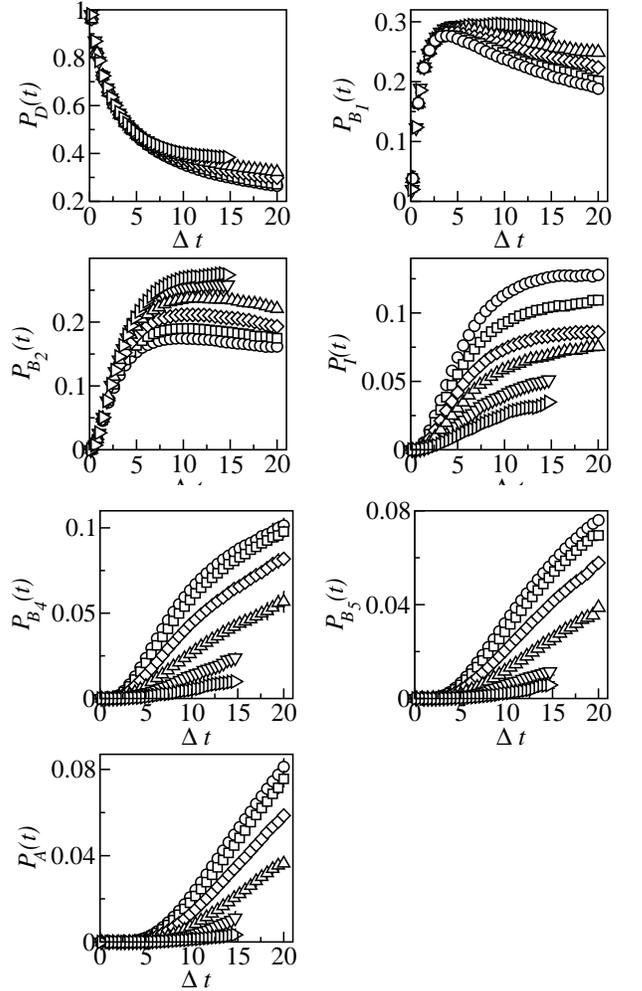

\epsfxsize=8cm \epsffile{fig_2a.eps} \epsfxsize=8cm
\epsffile{fig_2b.eps} \caption[]{\label{fig2a} Electronic
populations for $\alpha = 1/4$, $\omega_c/\Delta = 5$,
$\hbar\beta\Delta = 0.1$, $\epsilon_B/\Delta = 2.5$ and
$\epsilon_I/\Delta = 0, 2.5, 5, 7.5, 10$, and $12.5$
(circles, squares, diamonds, triangles up, triangles down,
and triangles right, respectively).}
\end{figure}
\begin{figure}
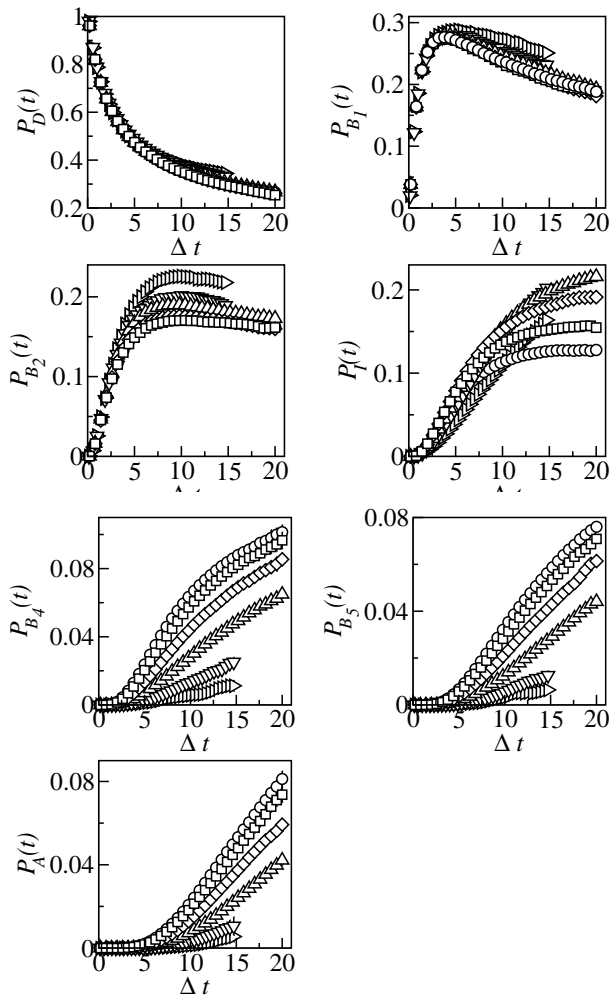

\epsfxsize=8cm \epsffile{fig_2c.eps} \epsfxsize=8cm
\epsffile{fig_2d.eps} \caption[]{\label{fig2b}Electronic
populations for $\alpha = 1/4$, $\omega_c/\Delta = 5$,
$\hbar\beta\Delta = 0.1$, $\epsilon_B/\Delta = 2.5$ and
$\epsilon_I/\Delta = 0, -2.5, -5, -7.5, -10$, and $-12.5$
(circles, squares, diamonds, triangles up, triangles down,
and triangles right, respectively).}
\end{figure}
The intersite rates were then obtained from our QMC data by the following
procedure: For increasing values of the transient time scale ${\tau}_{\rm
trans}$ starting at ${\tau}_{\rm trans}=0$ , the populations stemming from
solving Eq.~(\ref{differential rate equation}) with integration constants
specified by the QMC results for $P_A({\tau}_{\rm trans}), \dots,
P_D({\tau}_{\rm trans})$ were fitted to the QMC data for $P_A(t \ge {\tau}_{\rm
trans}), \dots, P_D(t \ge {\tau}_{\rm trans})$. This fixed the intersite rates
as functions of the transient time scale. Each true rate constant then was
extracted as the plateau value of its corresponding rate function with
$\tau_{\rm trans}$ as the actual transient time. The rates were fitted in
sequel rather than at once. First, $\Gamma_{DB}$ and $\Gamma_{B}$ were obtained
from a fit to a completely degenerated bridge, i.e.\ $\epsilon_I=0$, with
$\hbar\beta\Delta = 0.1$ and $\epsilon_B = 2.5\Delta$, yielding
\begin{eqnarray}
\Gamma_{DB}/\Delta &=& 0.298 \;, \nonumber\\
\Gamma_{BD}/\Delta &=& 0.383 \;, \nonumber\\
\Gamma_{B}/\Delta  &=& 0.370 \;,
\end{eqnarray}
which basically coincide with the golden rule values, see Ref.~[I],
Eq.~(22). For impurity bridges, $\epsilon_I \neq 0$, the bridge-impurity rate
$\Gamma_{BI}$ remains thus as the only parameter in the sequential population
dynamics (\ref{differential rate equation}). A subsequent fit yields
$\Gamma_{BI}$ as a function of the impurity offset $\epsilon_I$. The existence
of time independent intersite rates (plateau time) and the accuracy of this
procedure proofs in turn the applicability of the hopping model.

We start by presenting as an example in Figs.~\ref{fig2a},
\ref{fig2b} PIMC data for the population dynamics of a
bridge with $d = 7$. Our simulations included bridges with
$b = 3, 5, 7$ ($d = 5, 7, 9$) and gap energies between
$-12.5 \leq \epsilon_I/\Delta \leq 12.5$ in steps of
$2.5$. To obtain rate constants with sufficient accuracy,
the simulations had to be performed with exceptional small
statistical error and up to long times $\Delta\, t = 20$.
For example, for a typical coupling $\Delta=300$~cm$^{-1}$
this corresponds to simulation times up to 60~ps.
%
\begin{figure}
\epsfxsize=8cm \epsffile{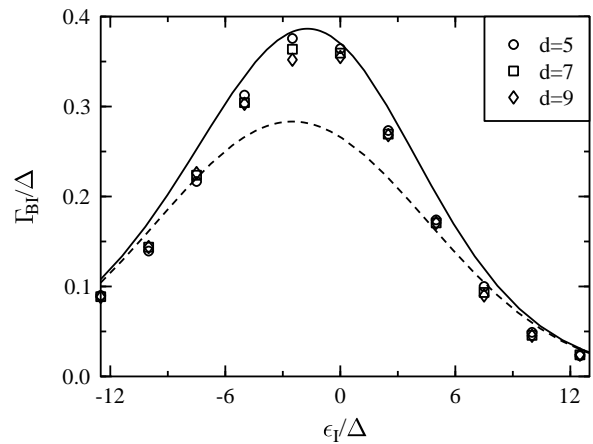} \caption[]{\label{fig3}
Bridge-impurity rate $\Gamma_{BI}$ as a function of the
impurity energy gap $\epsilon_I$. Rates extracted from
PIMC simulations (markers) for various chain length with
$d = 5, 7, 9$ are shown together with the results
according to the classical Marcus (dashed) and the quantum
mechanical golden rule (solid) expressions.}
\end{figure}
%
In Fig.~\ref{fig3} we display corresponding results for $\Gamma_{BI}$ as a
function of the energy gap $\epsilon_I$. Qualitatively, the numerical rates
follow the golden rule formula, but deviate, as expected, from the classical
Marcus expression [see Ref.~[I], Eq.~(21)].  Quantitatively, however, they are
smaller than the golden-rule results by about 10\% or less, a fact that can
partly be attributed to adiabatic effects like recrossings in the Landau-Zener
region since our bath is not strictly in the nonadiabatic limit where the
golden rule expression is valid. Rates for different bridge lengths differ by
less than 2\% outside the maximum and by about 5\% at the maximum where those
for shorter bridges exceed those of longer ones. In the classical limit, the
pronounced maximum of $\Gamma_{BI}(\epsilon_I)$ corresponds to an
activationless transfer at $\epsilon_I = -\Lambda_{\rm cl}$, while for
$\hbar\beta\omega_c \ge 1$ collective quantum effects in the bath promote
nuclear tunneling between adiabatic surfaces, shifting its location towards
smaller gap energies (cf.~fig.~\ref{fig3}). Note that $\Gamma_{BI} > \Gamma_B$
around the maximum. Deviations between rates for different $b$ are the largest
close to this maximum where adiabatic effects are expected to be the strongest
as the bath is directly located around the transition point.

The non-monotonic behavior of $\Gamma_{BI}(\epsilon_I)$ is already well-known
from Marcus theory since the activation barrier between bridge and impurity
state is a quadratic function of $\epsilon_I + \Lambda$, see Ref.~[I],
Eq.~(21).  Therefore, Marcus rates show perfect symmetry with respect to
$\epsilon_I = -\Lambda = -2.5\Delta$. The fact that acceptor populations in
Fig.~\ref{fig2a} and \ref{fig2b}, however, exhibit a symmetric behavior with
respect to $\epsilon_I = 0$, simply reflects the symmetry properties of the
total rate $\Gamma_{BI}(\epsilon_I) + \Gamma_{IB}(\epsilon_I) =
\Gamma_{BI}(\epsilon_I) + \Gamma_{BI}(-\epsilon_I)$. For long times, of course
this symmetry in the $P_A(t)$ is broken due to the different equilibrium
populations.
%
\begin{figure}
\vspace*{0cm} \epsfxsize=8cm \epsffile{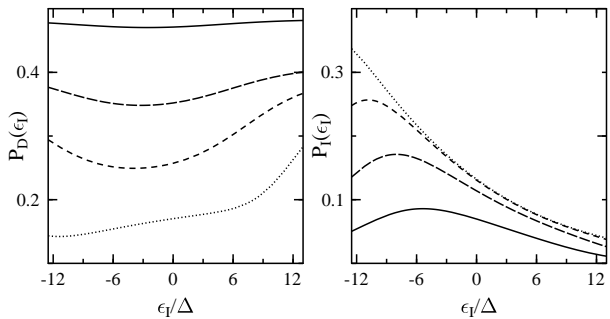}
\caption[]{\label{fig4} Donor population $P_D$ (left) and
impurity population $P_I$ (right) for $d=7$ at fixed times
vs.\ $\epsilon_I$. Left: $\Delta t=5$, (solid), 10
(long-dashed), 20 (short-dashed), 100 (dotted); right:
$\Delta t=5$, (solid), 10 (long-dashed), 20 (short-dashed),
50 (dotted).}
\end{figure}

Based on these results and Eq.~(\ref{differential rate equation}), we show in
Fig.~\ref{fig4} populations $P_D(t)$ and $P_I(t)$ for $b=5$ at fixed times but
for varying $\epsilon_I$. This gives deeper insight in how the final
equilibrium populations (\ref{equipops}) are established during the time
evolution. Namely, the impurity population tends to establish an exponential
dependence $\propto \exp(-\hbar\beta\epsilon_I)$ on a much shorter time scale
than the donor one displays a behavior $\propto \exp[\hbar\beta
(\epsilon_B+\epsilon_I)]$. We will see in detail below that this is a
characteristic facet of the DBA dynamics.

\subsection{Total transfer rate}
\label{Subsec:Total transfer rate}

With $\Gamma_{IB}$ at hand, the transfer dynamics according to
Eq.~(\ref{differential rate equation}) is known for arbitrary times. In
particular, the least non-vanishing eigenvalue of the rate matrix $A$ defines
the total transfer rate $\Gamma_T$ being typically, e.g.\ for longer bridges,
much smaller than the intersite rates.  This way, we get access to information
in the extreme long time range by extracting intersite rates from exact PIMC
simulations.
%
\begin{figure}
\epsfxsize=8cm \epsffile{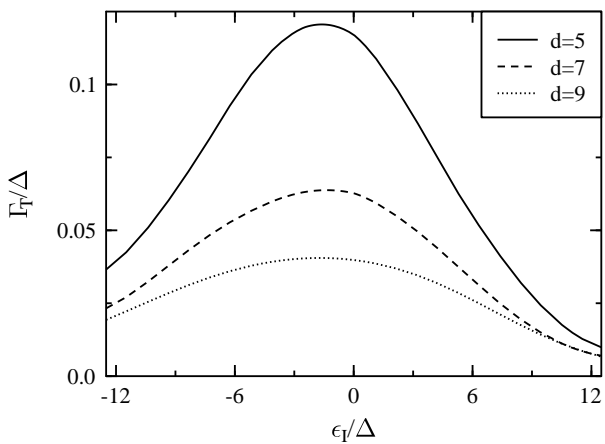} \caption[]{\label{fig5}
Total transfer rate $\Gamma_T$ according to the intersite
hopping rates extracted from the PIMC simulations and the
rate dynamics Eq.~(\ref{differential rate equation}) for
molecular chains with $d=5, 7, 9$ sites vs.\ the impurity
gap $\epsilon_I$.}
\end{figure}

Results are depicted in Fig.~\ref{fig5} for varying gap energy $\epsilon_I$ and
bridge lengths $b=3, 5, 7$. As discussed, the overall values of $\Gamma_T$ are
smaller by about a factor of 5 to 10 compared to the intersite hopping rates
meaning that the time scales for local dynamics are well separated from global
relaxation. Strikingly, there is a maximum at $\epsilon_I/\Delta\approx -1.5$
which is a signature of the activationless regime found in the $\Gamma_{BI}$
rate. Hence, a defect with an appropriate negative gap enhances the global
transport dynamics compared to the situation where it is absent. Of course, for
very large negative and positive gaps the chain is basically cut into pieces
and the sequential transfer to the acceptor tends to vanish.

\begin{figure}
\vspace*{0cm} \epsfxsize=8cm \epsffile{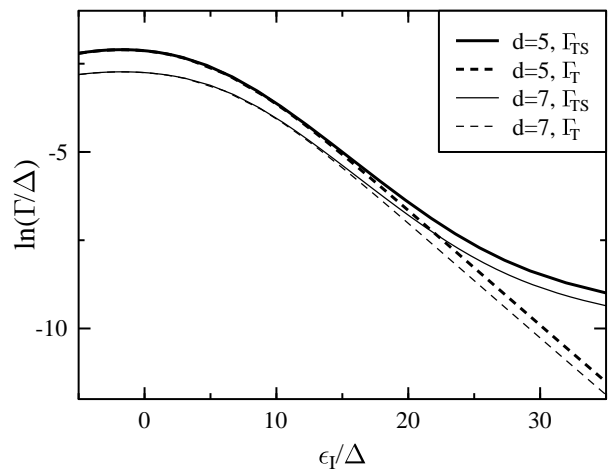}
\caption[]{\label{fig6} Total transfer rates without
($\Gamma_T$, dashed) and including superexchange across the
impurity site ($\Gamma_{TS}$, solid) vs.\ $\epsilon_I$.}
\end{figure}

In our previous studies, non-sequential electronic transport processes
(superexchange) played only a minor role. For the range of gap energies
considered in the PIMC simulations, the same holds true here. Since the rates
gained from the PIMC data are close to the golden rule rates, particularly for
higher/lower impurity sites, we analyse the situation for extreme gap energies
by additionally including direct transfer without populating the impurity,
i.e.\ from $|B_{(d+1)/2-1}\rangle$ to $|B_{(d+1)/2+1}\rangle$, into the rate
matrix $A$ via a superexchange rate $\Gamma_{SE}$. Knowing that superexchange
is relevant for high lying impurities only, we estimate $\Gamma_{SE}$ from a
quantum mechanical golden rule expression as well (forth order perturbation
theory in $\Delta$), see Eqs.~(23), (25) in Ref.~[I]. The corresponding
$\Gamma_{TS}(\epsilon_I)$ is shown together with $\Gamma_T$ in Fig.~\ref{fig6}
and verifies that for the total transfer, sequential hopping dominates by far
up to impurities with $\epsilon_I/\Delta \leq 20$.  Superexchange dominates for
larger gaps though, where the transport to the acceptor is almost frozen. In
this latter regime {\em electronic tunneling} through the impurity exceeds the
effect of {\em nuclear tunneling} in the sequential hopping mechanism.
%

\subsection{Bridge relaxation}
\label{Subsec:Bridge relaxation}

Even though the molecular bridge including the defect plays a key role in the
charge transfer from donor to acceptor, typically in experiments individual
bridge site populations are not accessible. Thus, the question arises whether
the bridge compound considered as a sub-unit of the entire molecular DBA
structure exhibits a characteristic dynamics on its own. For that purpose, we
look at the total bridge population $P_B=\sum_l\, P_{B_l}$ where the sum runs
over all bridge sites including the impurity. Clearly, due to normalization
$P_B = 1 - P_D - P_A$.
%
\begin{figure}
\epsfxsize=8cm \epsffile{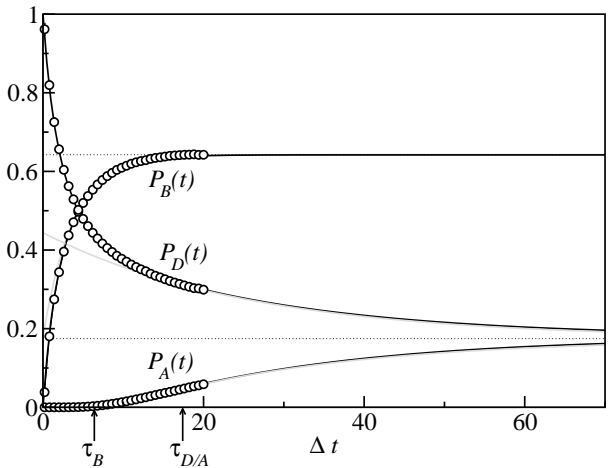} \caption[]{\label{fig7}
Time dependent populations of bridge $P_B(t)$, donor
$P_D(t)$, and acceptor $P_A(t)$ for a bridge with $b=5$
sites and an impurity gap $\epsilon_I/\Delta=5$. Dots
represent PIMC data, solid lines the full population
dynamics according to Eq.~(\ref{differential rate
equation}), and grey lines the asymptotic behavior
according to Eqs.~(\ref{asymbridge}) and (\ref{simple}),
see text. The dotted lines indicate the equilibrium
populations.}
\end{figure}
\begin{figure}
\epsfxsize=8cm \epsffile{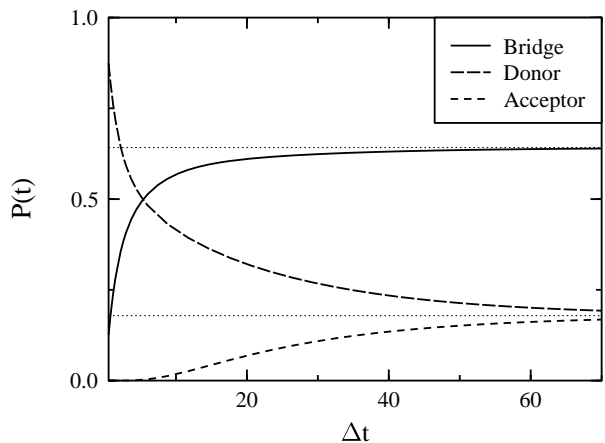} \caption[]{\label{fig8}
Time dependent populations of bridge $P_B(t)$, donor
$P_D(t)$, and acceptor $P_A(t)$ for a bridge with $b=5$
sites and an impurity gap $\epsilon_I/\Delta=5$ where the
impurity is located one site to the left of the center of
the bridge. Dotted lines indicate the respective
equilibrium values.}
\end{figure}

\begin{figure}
\vspace*{0.5cm} \epsfxsize=8cm \epsffile{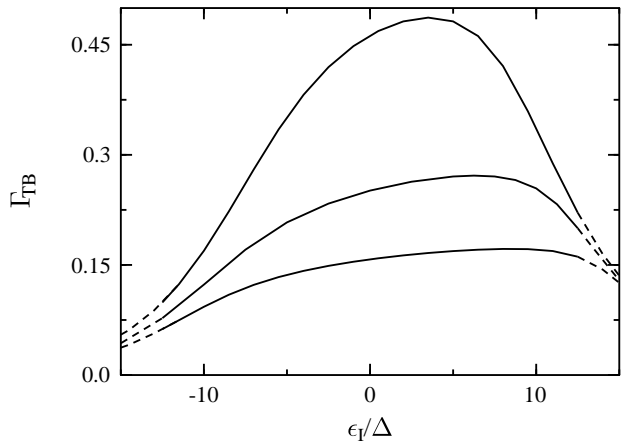}
\caption[]{\label{fig9} Bridge relaxation rate
$\Gamma_{TB}$ vs.\ $\epsilon_I$ for bridges with $b=3, 5,
7$ (from top to bottom), see text for details. The solid
lines are based on the PIMC data, while for
$|\epsilon_I/\Delta|>12.5$ the dashed lines are
extrapolations based on golden rule hopping rates to
better see the fall-off. }
\end{figure}

Upon closer inspection one finds that due to the symmetry
of the rate matrix in Eq.~(\ref{differential rate
equation}), every second term in $P_B(t)$ drops out
exactly so that its asymptotic dynamics is governed by the
second least eigenvalue of $A$, denoted by $\Gamma_{TB}$,
instead of $\Gamma_T$. For times beyond a certain
threshold $\tau_B\ll 1/\Gamma_T$, the bridge saturation can
thus be captured by
\begin{equation}
P_B(t)=P_B^\infty+[P_B(\tau_B)-P_B^\infty]\, {\rm
e}^{-\Gamma_{TB} (t-\tau_B)}\, \label{asymbridge}
\end{equation}
where $\Gamma_{TB} \gg \Gamma_T$; accordingly, even if the
intersite rates in $A$ are not known, $\Gamma_{TB}$ can
still be found by studying the saturation limit of the
time dependent expression
$\Gamma_{TB}(t)=\dot{P}_B(t)/|P_B(t)-P_B^\infty|$. This
theoretical prediction from the hopping model is
completely verified by the simulations as shown in
Fig.~\ref{fig7}, where $P_B(t)$ for fixed gap
$\epsilon_I/\Delta=5$ is depicted as a function of time.
Apparently, it tends to saturate on a considerably smaller
time scale than the total chain relaxes towards thermal
equilibrium. An explicit value for the transient time
$\tau_B$ is gained from the condition that the actual
$P_B(t)$ and its asymptotics (\ref{asymbridge}) differ by
less than 1\%. This leads to $\Delta \tau_B=6.4$ which is
considerably below the simulation time. As seen in
Fig.~\ref{fig9}, the bridge relaxation rate takes its
maximum not at negative, but rather positive gap energies,
where the overall transfer is slowed down. In this case,
the defect facilitates as a scattering center first
relaxation to its left side and then to its right one by
blocking the flux to the acceptor site. Note again that
the asymptotic dynamics (\ref{asymbridge}) of the total
compound does not mean that the {\em individual} bridge
site populations have approached their equilibrium values,
too.

Now, for very long times $t\gg 1/\Gamma_{TB}$ one has $\dot{P}_D(t)=
-\dot{P}_A(t)$ so that then the transport across the DBA complex can be
described within an effective two state model with a transfer rate being
identical to the total transfer rate $\Gamma_T$ extracted above, i.e.,
\begin{eqnarray} \label{simple}
P_D(t) &=& P_D^\infty + [P_D(\tau_{D/A})-P_D^\infty]\, {\rm
e}^{-\Gamma_T (t-\tau_{D/A})}
\nonumber\\
P_A(t) &=& P_D^\infty - [P_D(\tau_{D/A})-P_D^\infty]\, {\rm
e}^{-\Gamma_T (t-\tau_{D/A})}\,.
\end{eqnarray}
Here, besides $\Gamma_T$ only information about the donor enters. Further,
$\Delta \tau_{D/A}=17.4$ (extracted similar as described above for $\tau_b$)
obeys $\tau_{D/A}\gg 1/\Gamma_{TB}$ and lies also well within the simulation
time window, thus verifying that the PIMC simulations cover the entire relevant
time range.  Everything beyond this range is in a sense "simple" and completely
determined by the information retrieved from the numerical data. The comparison
between Figs.~\ref{fig5} and \ref{fig9} reveals that indeed $\Gamma_{TB}\gg
\Gamma_T$.

While this strong time separation is no longer true for a defect located off
center, the total bridge population comes still significantly closer to its
equilibrium value on a time scale much below $1/\Gamma_T$ than $P_D$ and $P_A$,
see Fig.~\ref{fig8}. The remaining deviation falls off with $\Gamma_T$, but
practically Eq.~(\ref{simple}) still gives an accurate description of the D-A
transfer at long times. Hence, the underlying scenario for the DBA transfer is
this: On an intermediate time scale $1/\Gamma_{TB}$ a local steady state for
the total bridge population is established, which then acts for larger times
merely as an electronic reservoir mediating between D and A. In turn, the
comparison between the dynamics of $P_A(t)+P_D(t)$ and the individual
populations $P_D(t), P_A(t)$ would allow experimentally to gain information
about the symmetry of the bridge/defect structure. In this context it is
worthwhile to mention that a symmetric situation treated here numerically is
not just an idealized toy model but can actually be realized in experiments,
see Ref.~\cite{weber1}.

The above findings on the time scale separation allow for a simple recipe to
successively extract rate constants from the donor and acceptor dynamics
only. Namely, in a first step $\Gamma_T$ is obtained from the asymptotic
behavior of $P_D(t)$ and $P_A(t)$. Then, for a symmetric system $P_D(t)+P_A(t)$
allows to obtain the second least rate constant $\Gamma_{TB}$. Further, by
subtracting the dynamics corresponding to these rates from $P_D(t)$ and
$P_A(t)$ the third least eigenvalue shows up etc. Practically, this procedure
works for the rate constants governing the dynamics from moderate to long times
for which typically a sufficient separation of time scales is guaranteed.
%

\section{Summary and Conclusions}

Based on a recently improved real-time QMC scheme to treat numerically exact
electronic transfer dynamics in condensed phase, we analysed the impact of an
impurity on transport across molecular chains. Experimentally, this is not only
the typical situation in native structures, but advanced molecular synthesis
even allows to assemble corresponding chains in a controlled way. We restricted
our study to the case of an impurity site with an energy gap to its neighboring
sites, which can be seen as the archetypical scenario as it provides e.g.\ also
an effective description of a defect in the electronic intersite coupling.

Our main focus was the total transfer rate corresponding to the time scale for
global relaxation towards thermal equilibrium. Since for longer bridges this
time scale is in general considerably larger than the largest times accessible
in real-time PIMC simulations, we used the latter ones to extract, beyond a
transient time, intersite hopping rates between the electronic sites. This
information was then used in a sequential transfer model, verified by the PIMC
data, to capture the extreme long time dynamics.

Roughly, two dynamical ranges can be distinguished in the charge transfer: In a
first step, the total bridge population approaches (symmetric bridge/defect) or
comes close to (asymmetric bridge/defect) its value for thermal equilibrium
facilitated by the presence of an impurity, while in a second step overall
relaxation sets in associated with a characteristic unique time scale, the
total transfer rate.  Even though the total bridge population tends to saturate
in the first range, intersite relaxation between individual sites still takes
place.  Nevertheless, the bridge compound including the impurity can basically
be treated as a thermal reservoir mediating the transfer from donor to
acceptor. The extent to which a time scale separation between bridge and total
relaxation can be observed is directly related to the bridge/defect
symmetry. The total transfer rate exhibits a pronounced maximum at negative
values of the impurity energy gap, thus reflecting the enhancement of the local
bridge-impurity rate in or near the activationless regime.  In principle, a
controlled doping with a defect site may therefore accelerate the transfer
across a degenerated chain. As in our previous study, superexchange across the
impurity plays a dominant role for extremely large gaps where the transport
comes almost to rest. Our analysis may give qualitative and quantitative
insight into experiments similar to the one reported in Ref.~\cite{weber1}
where a molecular chain contacted with leads was designed to include a defect
at its center.

\acknowledgements
We acknowledge financial support from the DFG through
Grant No. AN336-1  and the Landesstiftung
Baden-W{\"u}rttemberg gGmbH. J.A.\ is a Heisenberg fellow
of the DFG.



\begin{thebibliography}{99}
%
\bibitem{ratner} J. Jortner and M.A. Ratner (eds.), {\em Molecular
Electronics}, (Blackwell Sci., Oxford, 1997).
%
\bibitem{jortner} J. Jortner and M. Bixon, eds., Adv. Chem. Phys. {\bf
106}, 107 (1999).
%
\bibitem{dekker} C. Dekker and M.A. Ratner, Phys. World {\bf 14}, 29
(2001).
%
\bibitem{nitzan2} A. Nitzan, Ann. Rev. Phys. Chem. {\bf 52}, 681 (2001).
%
\bibitem{hanggi} P. H{\"a}nggi and M.A. Ratner (eds.), Chem. Phys. {\bf
281}, (2002).
%
\bibitem{saclay} C. Kergueris, J.-P. Bourgoin, S. Palacin, D. Esteve, C.
Urbina, M. Magoga, and C. Joachim, Phys. Rev. B {\bf 59},
12505 (1999).
%
\bibitem{weber} J. Reichert, R. Ochs, H.B. Weber, M. Mayor, and H.v. L\"ohneysen,
Appl. Phys. Lett. {\bf 82}, 4137 (2003).
%
\bibitem{marcus56} R.A. Marcus, J. Chem. Phys. {\bf 24}, 966 (1956).
%
\bibitem{marcus85}
R.A. Marcus and N. Sutin, Biochim. Biophys. Acta {\bf
811}, 265 (1985).
%
\bibitem{weiss}
U. Weiss, {\sl Quantum Dissipative Systems}, Series in
Modern Condensed Matter Physics, Vol. 2 (World Scientific,
Singapore, 1998).
%
\bibitem{friesner} A.K. Felts, W.T. Pollard, and R.A. Friesner, J. Phys.
Chem. {\bf 99}, 2929 (1995).
%
\bibitem{mukamel} A. Okada, V. Chernyak, and S. Mukamel, J. Phys. Chem. A
{\bf 102}, 1241 (1997).
%
\bibitem{nitzan0} W.B. Davis, M.R. Wasielewski, M. Ratner, V. Mujica, and
A. Nitzan, J. Phys. Chem. {\bf 101}, 6158(1997).
%
\bibitem{nitzan1} D. Segal, A.Nitzan, W.B. Davis, M.R. Wasielewski, and
M.A. Ratner, J. Phys. Chem. {\bf 104}, 3817 (2000).
%
\bibitem{last_paper} L. M{\"u}hlbacher, J. Ankerhold, and
C. Escher, J. Chem. Phys. {\bf 121}, 12696 (2004).
%
\bibitem{egger_mak}
C.H. Mak and R. Egger, Adv. Chem. Phys. {\bf 93},39 (1996).
%
\bibitem{joachim} M. Magoga and C. Joachim, Phys. Rev. B {\bf 56}, 4722
(1997).
%
%
\bibitem{ratner_nature} W.B. Davis, W.A. Svec, M.A. Ratner, and M.R.
Wasielewski, Nature {\bf 396}, 60 (1998).
%
\bibitem{giese} B. Giese, J. Amaudrut, A.-K. K{\"o}hler, M. Spormann,
  and S. Wessley, Nature {\bf 412}, 318 (2001).
%
\bibitem{weber1} M. Mayor, C. von H{\"a}nisch, H.B. Weber,
J. reichert, D. Beckmann, Angew. Chemie {\bf 41}, 1183
(2002).
%
\bibitem{chandler}
D. Chandler, in {\sl Liquids, Freezing and the Glass
Transition}, ed.~by D. Levesque, J.P. Hansen. and J.
Zinn-Justin (Elsevier Science, North Holland, 1991), Les
Houches 51, Part 1.
%
\bibitem{song}
X. Song and A.A. Stuchebrukhov, J. Chem. Phys. {\bf 99},
969 (1993).
%
%
\bibitem{mlb03}
L. M{\"u}hlbacher and R. Egger, J. Chem. Phys. {\bf 118},
179 (2003).
%
\bibitem{feynman}
R.P. Feynman and F.L. Vernon, Ann. Phys. (N.Y.) {\bf 24},
118 (1963).
%
\bibitem{sign-problem}
 {\em Quantum Monte Carlo
Methods in Condensed Matter Physics}, M. Suzuki (ed.)
(World Scientific, Singapore, 1993), and references
therein.
%
\bibitem{nica} R. Egger, C.H. Mak, and U. Weiss, Phys. Rev. E {\bf 50},
(655)(R) (1994).
%

\end{thebibliography}
\end{document}